\documentclass[%
 aip,
 amsmath,amssymb,
 reprint,%
]{revtex4-1}

\usepackage{graphicx}
\usepackage{dcolumn}
\usepackage{bm}

\usepackage[utf8]{inputenc}
\usepackage[T1]{fontenc}
\usepackage{mathptmx}
\usepackage{etoolbox}

\makeatletter
\def\@email#1#2{%
 \endgroup
 \patchcmd{\titleblock@produce}
  {\frontmatter@RRAPformat}
  {\frontmatter@RRAPformat{\produce@RRAP{*#1\href{mailto:#2}{#2}}}\frontmatter@RRAPformat}
  {}{}
}%
\makeatother
\begin{document}

\preprint{AIP/123-QED}

\title[Impact of pulse exposure on chimera state]{Impact of pulse exposure on chimera state in ensemble of FitzHugh-Nagumo systems}
\author{E. Rybalova}%
 \affiliation{Radiophysics and Nonlinear Dynamics Departament, Institute of Physics, Saratov State University, Astrakhanskaya str. 83, Saratov 410012, Russia}%
\author{N. Semenova}%
 \email{semenovani@sgu.ru}
 \affiliation{Radiophysics and Nonlinear Dynamics Departament, Institute of Physics, Saratov State University, Astrakhanskaya str. 83, Saratov 410012, Russia}%

\date{\today}

\begin{abstract}
In this article we consider the influence of a periodic sequence of Gaussian pulses on a chimera state in a ring of coupled FitzHugh-Nagumo systems. We found that on the way to complete spatial synchronization one can observe a  number of variations of chimera states that are not typical for the parameter range under consideration. For example, the following modes were found: breathing chimera, chimera with intermittency in the incoherent part, traveling chimera with strong intermittency, and others. For comparison, here we also consider the impact of a harmonic influence on the same chimera, and to preserve the generality of the conclusions, we compare the regimes caused by both a purely positive harmonic influence and a positive-negative one.\end{abstract}


\maketitle

\begin{quotation}
In ensembles of coupled elements, various complex spatiotemporal regimes can arise, including chimera states. These regimes represent the coexistence of coherent and incoherent clusters, and have been observed not only in experiments but also in real systems of different natures. One notable example is the dynamics of neurons in the brain during seizures of epilepsy and Parkinson's disease. This highlights the importance of understanding and managing such dynamic regimes. Since neural activity involves impulse dynamics, it is crucial to study the stability of modes under this type of external influence. This work focuses on investigating the influence of impulse signals on an ensemble of nonlocally coupled FitzHugh-Nagumo neurons, in which chimeras can occur.
\end{quotation}

\section{Introduction}\label{sec:intro}
In complex multicomponent systems with nonlinear subsystems, different spatiotemporal regimes can be established. These regimes can have either a favorable or destructive effect on the systems. The simplest and most common regimes include synchronization of subsystems in amplitude, frequency, and phase, as well as complete desynchronization of subsystems. In 2002, Y. Kuramoto and D. Battogtokh demonstrated that in an ensemble of nonlocally coupled phase oscillators, it is possible to establish a spatiotemporal regime characterized by the coexistence of a cluster with incoherent dynamics of elements and a cluster with coherent dynamics of elements~\cite{Kuramoto:2002uu}. Two years later, D.M. Abrams and S.H. Strogatz called this dynamic regime a ``chimera state''~\cite{Abrams:2004vx}. This term sparked additional research interest in this phenomenon, and this interest remains active to this day. The importance of studying this phenomenon is evidenced by its occurrence not only in computer modeling~\cite{Omelchenko:2011uc,Omelchenko2013,Panaggio:2015uu,Maistrenko:2015vq,Semenova:2015tt,Scholl:2016vm,Kemeth:2016vc,Ulonska:2016tx,Shepelev:2017uy,Bogomolov:2017wq,Bukh:2017vp,Zakharova:2020uu} but also in real systems of various nature~\cite{Menck:2014tw,Motter:2013vz,Wang:2016wb,Gonzalez-Avella:2014vg,Hong:2010wf}, including neural networks~\cite{Levy:2000aa,Rattenborg:2000aa, Andrzejak:2016aa,Lainscsek:2019aa,Huo:2021aa,Bressloff:2011aa,Zhang:1996aa,Bansal:2019wy,Scholl:2022aa}.

Neural networks contain a large number of neurons with complex connections, which makes the direct study of such systems difficult. Currently, neural ensembles are initially studied using model systems in computer experiments. In this case, various models of neurons are used, which, to some extent, replicate the processes occurring in real neurons. These include the Hodgkin-Huxley model\cite{Hodgkin:1952ab}, FitzHugh--Nagumo model\cite{Fitzhugh:1960aa,Nagumo:1962vi}, Morris--Lecar model\cite{Morris:1981aa}, Hindmarsh--Rose model\cite{Hindmarsh:1984aa} and their various modifications. One of the most commonly used neuron models is the FitzHugh--Nagumo oscillator, which is a simplified modification of the Hodgkin--Huxley model. This model not only reproduces the main effects observed in neurons but also consists of only two equations and a few control parameters, making mathematical and computer studies of the dynamics of this model easier.

Studies have shown that ensembles of nonlocally coupled FitzHugh-Nagumo oscillators can exhibit chimera states when partial elements are in both oscillatory and excitable modes. In an ensemble of excitable FitzHugh-Nagumo oscillators influenced by noise, coherence-resonance chimeras~\cite{Semenova:2016aa,Zakharova:2017wx,Semenova:2020aa} and self-induced-stochastic-resonance chimeras~\cite{Zhu:2023aa,Khatun:2023aa} have been discovered. However, previously chimera states were found in ensembles of FitzHugh-Nagumo models operating in the oscillatory regime. This was achieved through cross coupling between an activator and an inhibitor~\cite{Omelchenko2013,Omelchenko:2015uu,Ramadoss:2021aa}

The inherent nature of neurons and neural ensembles is characterized by impulsivity, making impulse action a longstanding method for operating and controlling neural activity. An illustrative example of this is the use of electrical stimulation via deep brain stimulation (DBS) electrodes, which are implanted to treat motor problems in Parkinson's disease and dystonia~\cite{Benabid:1991aa,Armstrong:2020aa,Hale:2020aa,Fan:2021aa,Hariz:2022aa}. Additionally, DBS is employed in the treatment of various neurological disorders such as Tourette syndrome~\cite{Xu:2020aa,Szejko:2022aa}, depression~\cite{Figee:2022aa,Wu:2021aa}, and even in the treatment of epilepsy~\cite{Vetkas:2022aa,Fisher:2023aa}, which serves as a prime example of a chimera state in the brain.

In this paper, we consider the effect of pulse exposure on a chimera state in a ring of coupled FitzHugh-Nagumo systems. This is the first work in a series of works in which we plan to study the destruction of spatio-temporal regimes that arise in ensembles of neuron models under external influence. Here we are mainly focused on periodic pulse exposure, but later this list will be expanded to noisy pulse action and L\'evy noise.

\section{System under study}\label{sec:system}
\subsection{Ring of FHN systems}\label{sec:system_FHN}
Here we consider the ring of nonlocally coupled FHN systems under external influence $I(t)$. The autonomous part of system is set in the same way as was in Ref.~\cite{Omelchenko2013}: 
\begin{equation}\label{eq:FHN}
\begin{array}{c}
\varepsilon \frac{du_i}{dt}=u_i - \frac{u_i^3}{3}-v_i+ I(t)+ \\
\frac{\sigma}{2R}\sum\limits^{i+R}_{j=i-R} [b_{uu}(u_j-u_i) + b_{uv}(v_j-v_i)] , \\
\frac{dv_i}{dt}=u_i+a_i+\\ 
\frac{\sigma}{2R}\sum\limits^{i+R}_{j=i-R}[b_{vu}(u_j-u_i)+b_{vv}(v_j-v_i)],
\end{array}
\end{equation}
where $u_i$ and $v_i$ are the activator and inhibitor variables, respectively, the oscillator's number is set by $i=1,...,N$ with $N=500$ being the total number of elements in the network with periodic boundary conditions. A small parameter $\varepsilon>0$ sets the time scale separation of fast activator and slow inhibitor variables, while $a$ defines the excitability threshold. For each partial FHN system, it determines whether the system is in the excitable ($|a|>1$), or oscillatory ($|a|<1$) regime. 

The parameter $R$ indicates the number of nearest neighbours in each ring direction coupled with $i$th element. For convenience, we additionally introduce the coupling range set by $r=R/N$ . The strength of the coupling is characterized by $\sigma$. For our simulations we use initial conditions randomly distributed on circle $u^2+v^2\le 2^2$.

Equation.~(\ref{eq:FHN}) contains not only direct, but also cross couplings between activator ($u$) and inhibitor ($v$) variables, which can be modeled by a rotational coupling matrix \cite{Omelchenko2013} to reduce the number of parameters: 
\begin{eqnarray}
B=\left(
\begin{array}{cc}
b_{uu} & b_{uv}\\
b_{vu} & b_{vv}
\end{array}\right)
=\left(
\begin{array}{cc}
\cos\phi & \sin\phi\\
-\sin\phi & \cos\phi
\end{array}\right)
\end{eqnarray}
where $\phi\in[-\pi;\pi)$ is the parameter allowing to simultaneously control the contribution of all four connection types in (\ref{eq:FHN}). 

\begin{figure}[h]
\includegraphics[width=\linewidth]{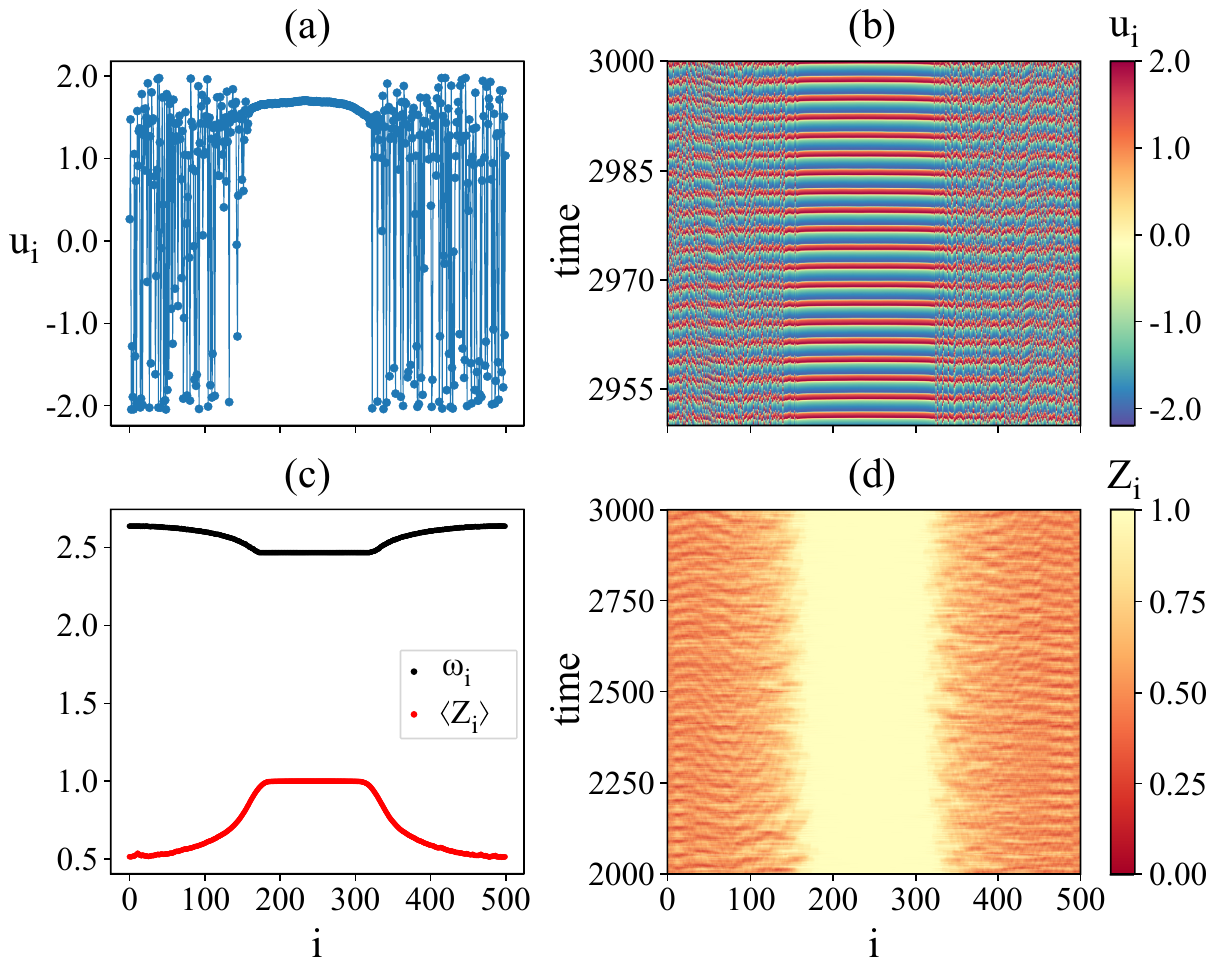}
\caption{\label{fig:chimera_ome13} Chimera state observed in autonomous system (\ref{eq:FHN}) illustrated by the instantaneous snapshot (a) and space-time plot (b) of variable $u$, space-time plot of local order parameter (d) and averaged local order parameter with mean phase velocity given in panel (c). Parameters: $\varepsilon=0.05$, $a=0.5$, $\phi=\pi/2-0.1$, $r=0.35$, $\sigma=0.2$, $I(t)=0$.}
\end{figure}

In this study, we are mainly focused on chimera state regime arising when partial elements demonstrate periodic spiking. Therefore, here we use the same parameters as was in Ref.~\cite{Omelchenko2013} for chimera state $\varepsilon=0.05$, $a=0.5$, $\phi=\pi/2-0.1$, $r=0.35$, $\sigma=0.2$ fixed throughout the article. One isolated FHN system demonstrates oscillatory regime and periodic spiking behaviour for these $a$ and $\varepsilon$ values. At the same time, the network parameters $r$, $\sigma$ and $\phi$ lead to chimera state with one incoherent and one coherent domains. Every partial element of the network demonstrates spikes, and in coherent domain these spikes are periodic. The regime typical for considered autonomous system ($I(t)=0$) is given in Fig.~\ref{fig:chimera_ome13}.

Now, after introduction of the system, let us consider ways to visualize various modes. On the one hand, we can consider just one of the variables, for example the variable $u$, and visualize it via the instantaneous snapshot of this variable in space (Fig.~\ref{fig:chimera_ome13}(a)) and its spatiotemporal evolution using the space-time plot (Fig.~\ref{fig:chimera_ome13}(b)).  

On the other hand, it is impossible to analyze these images for a long time, so in addition to this we will also use such statistical characteristics as the local order parameter $Z_i$ and the mean phase velocity $\omega_i$. The local order parameter $Z_i = |\frac{1}{2\delta} \sum\limits_{|j-i|\le\delta} e^{i\Theta_j}|$ is a real number in the range from 0 to 1, where 1 corresponds to coherent behavior in a certain area, while 0 to incoherent behavior. Here $\Theta_i = \arctan(v_i/u_i)$ is the geometric phase of $i$th FHN unit. The spatiotemporal evolution of local order parameter is given in Fig.~\ref{fig:chimera_ome13}(d). The absence of fluctuations in the local order parameter allows to analyze the state of the system over a long time interval, and get an information about the stationarity or non-stationarity of the coherent and incoherent domains. Figure~\ref{fig:chimera_ome13}(c) shows the local order parameter averaged throughout all integration time $\langle Z_i\rangle$ (red points). The mean phase velocity (black points in Fig.~\ref{fig:chimera_ome13}(c)) is calculated as the number of spikes $M_i$ divided by the integration time $w_i=2\pi M_i/\Delta T$. It is calculated separately for each $i$th oscillator. However, for analyzing the entire system regime, we use the local order parameter and the mean phase velocity, which are averaged over space and denoted as $\langle Z\rangle$ and $\langle \omega\rangle$, respectively.

\subsection{Pulse exposure}\label{sec:system_pulses}
In order to introduce a pulse exposure and maintain the integrability of the system, here we consider Gaussian pulses shaped as a Gaussian function \cite{Chomycz2009,Bauer1984}: 
\begin{equation}\label{eq:one_pulse}
P(t)=A\cdot\exp\Big(-\frac{t^2}{2\Delta t^2}\Big),
\end{equation}
where $A$ is the amplitude of pulse, parameter $\Delta t$ controls the width of Gaussian pulse. Its smaller values make the pulse narrower. At the same time, it is important to ensure that when integrating a system with pulse exposure, the integration step is sufficient to adequately simulate pulses, so the pulses should not be made too narrow.

In this paper we are mainly interested in pulse-periodic exposure. Equation~\ref{eq:one_pulse} describes only one impulse. In order to make the influence periodic, it is necessary to replace $t$ with a periodic function of $t$:
\begin{equation}\label{eq:periodic_pulses}
I(t)=A\cdot\exp\Big(-\frac{1}{2 s} \sin^2\frac{w t}{2}\Big).
\end{equation}
This is how the external pulse-periodic exposure $I(t)$ is set in (\ref{eq:FHN}). It describes periodic pulses with amplitude $A$, frequency $w$ and pulse width controlled by the parameter $s$. Further, we will consider the impact of external influence depending on its amplitude $A$ and period $T=2\pi/w$.

\section{Results}
\subsection{Regimes induced by periodic Gaussian pulses}

\begin{figure}[h]
\includegraphics[width=\linewidth]{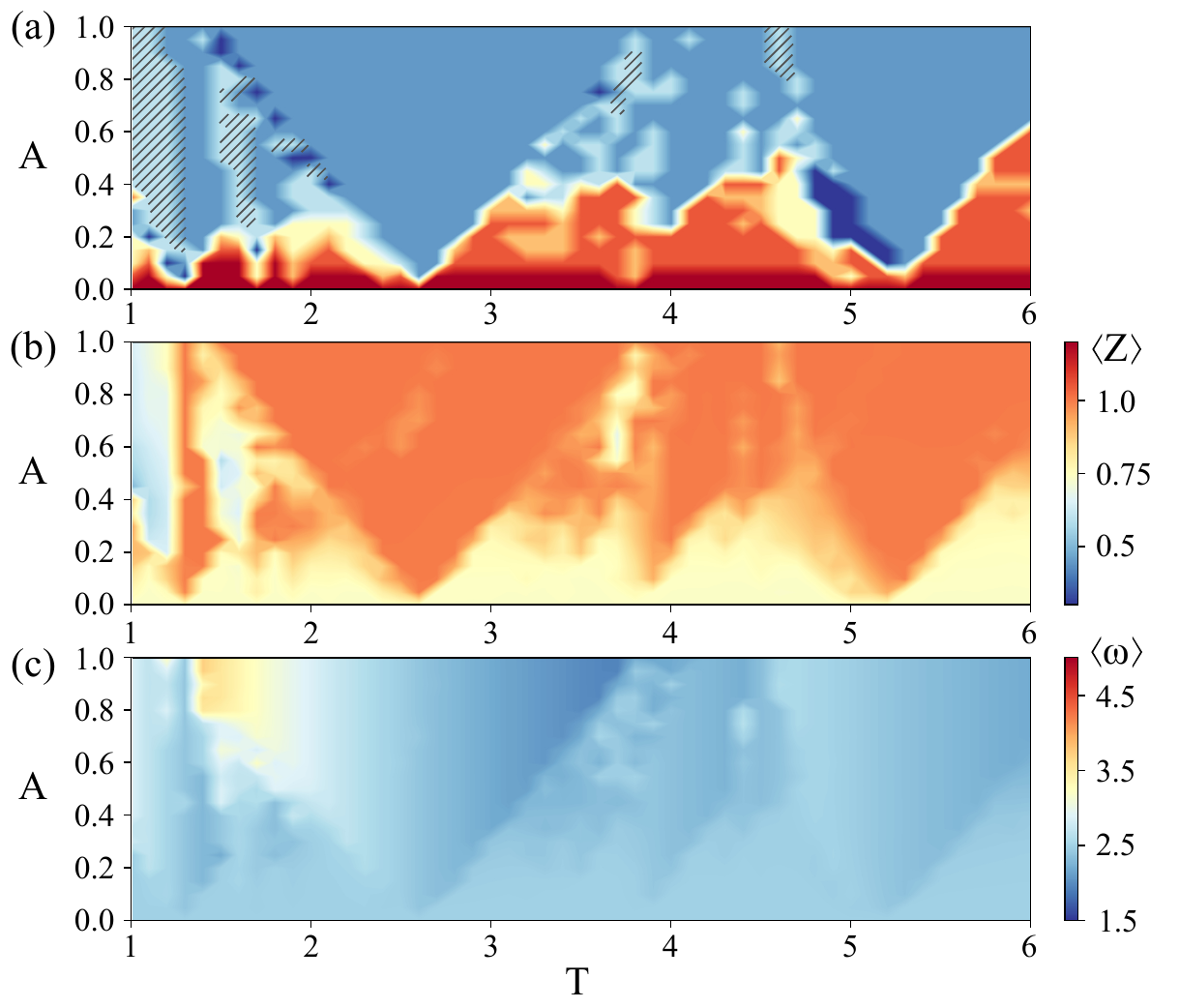}
\caption{\label{fig:maps_pulse} Impact of periodic Gaussian pulses on chimera state regime. Figure contains the map of regimes (a) in the parameter plane $(T, A)$, mean local order parameter $\langle Z\rangle$ averaged over the integration time and space (b), and mean phase velocity $\langle \omega\rangle$ averaged in the same way (c). Parameters: $s=0.1$, the rest parameters are the same as in Fig.~\ref{fig:chimera_ome13}.}
\end{figure}

In this section we consider the impact of periodic Gaussian pulses (\ref{eq:periodic_pulses}) on the ensemble of FHN systems (\ref{eq:FHN}) in chimera state regime. Figure \ref{fig:maps_pulse}(a) contains the map of regimes in the parameter plane of external influence period and amplitude. This map of regimes was prepared for four initial conditions. Half of the initial conditions were set randomly, and the other half were chimera states already stabilized in the system. The color scheme of the map Fig.~\ref{fig:maps_pulse} will be discussed further. In addition to the map of regimes, Fig.~\ref{fig:maps_pulse} contains also averaged local order parameter $\langle Z\rangle$ (b) and mean phase velocity (c) averaged throughout all integration time $t=3000$.

With a weak external influence $A\lessapprox 0.1$, the system demonstrates the original chimera state, marked in red on the map Fig.~\ref{fig:maps_pulse}(a). When the amplitude of influence becomes strong, complete synchronization is observed. The regions of complete spatial synchronization are show in middle-blue color in Fig.~\ref{fig:maps_pulse}(a) and look like Arnold tongues. They correspond to $\langle Z\rangle\approx 1$ in Fig.\ref{fig:maps_pulse}(b). As can be seen from the averaged mean phase velocity (c), the frequency of oscillations in ensemble are controlled by the frequency of external influence in these areas.

\begin{figure*}[t!]
\includegraphics[width=\linewidth]{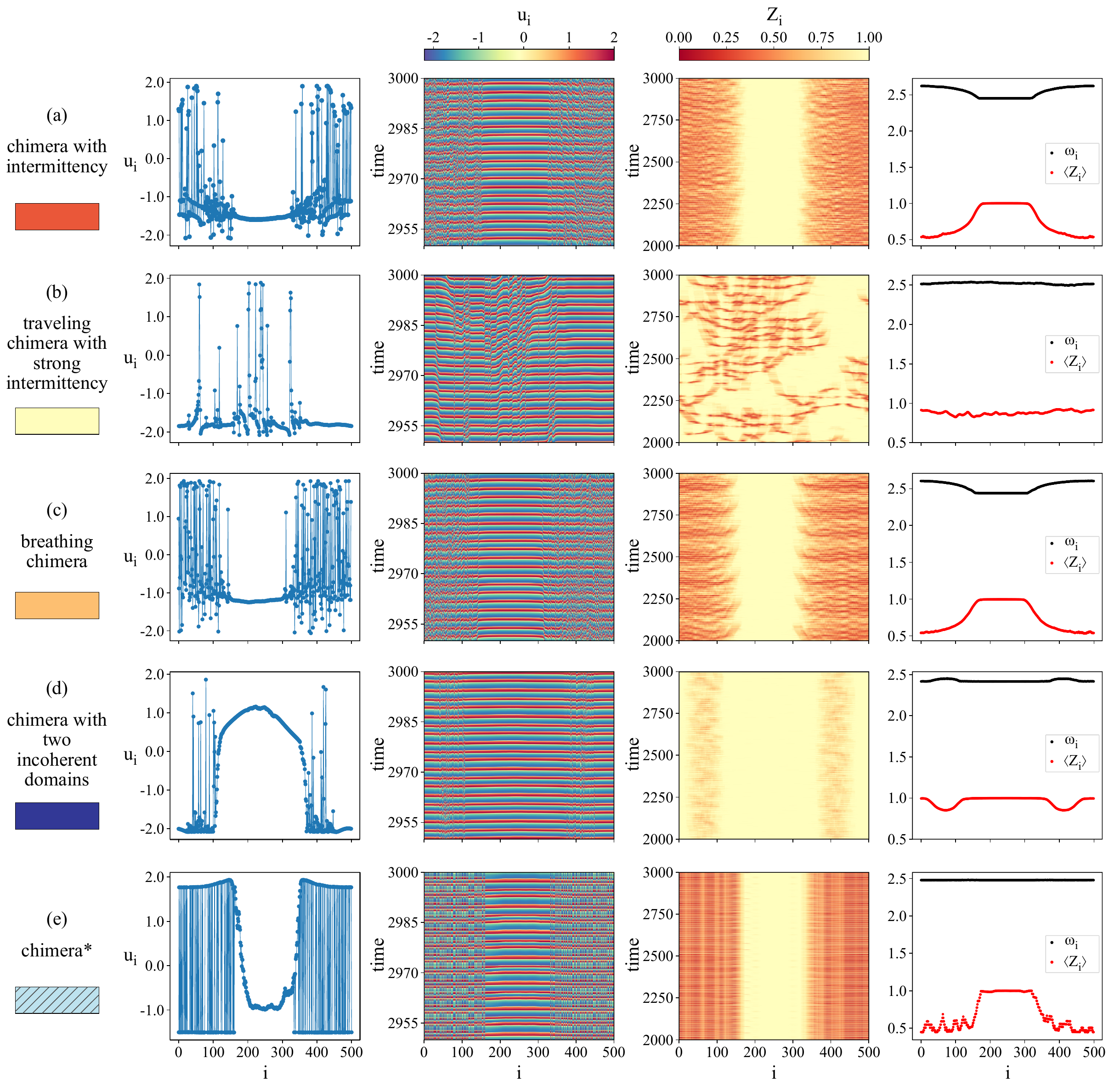}
\caption{\label{fig:zoo} Different regimes observed between original chimera state and complete synchronization in the case of periodic-pulse exposure. The first column of fragments shows the snapshots of variables $u_i$ at time $t=3000$, while their space-time plots are given in panels of the second column. The third column shows the space-time plots of the local-order parameter. The fourth column contains local order parameter and mean phase velocity averaged over time. Parameters: $A=0.15$, $T=3.5$ (a), $A=0.15$, $T=4.9$ (b), $A=0.15$, $T=3.2$ (c), $0.3$, $T=3.9$ (d), $A=0.8$, $T=3.8$ (e). The parameter $s=0.1$, while the rest parameters are the same as in Fig.~\ref{fig:chimera_ome13}.}
\end{figure*}

There are several modes observed between these two ones. They are given in Fig.~\ref{fig:zoo} using the same scheme of panels as was for original chimera in Fig.~\ref{fig:chimera_ome13}. Below we will describe each of these regimes separately.

There are two often found regimes between original chimera (Fig.~\ref{fig:chimera_ome13}) and complete spatial synchronization. The first one looks like original chimera state but there are chaotically appeared and disappeared small coherent domains inside the large incoherent domain (see Fig.~\ref{fig:zoo}(a) and compare incoherent clusters here and in Fig.~\ref{fig:chimera_ome13}). On the space-time plot of variable $u_i$ and local order parameter $Z_i$ (2nd and 3rd subpanels in Fig.~\ref{fig:zoo}(a)) this regime looks like original chimera state, but when analysing the snapshot (first subpanel in Fig.~\ref{fig:zoo}(a)) one can see, that incoherent domain is not so incoherent as was previously. We will call this regime ``chimera with intermittency''. This mode is stable and does not disappear with a long integration time. The position of the incoherent domain remains in the same place (3rd and 4th subpanels in Fig.~\ref{fig:zoo}(a)). This regime can be observed in the dark-orange area of the map Fig.~\ref{fig:maps_pulse}(a).

When parameters $A$ and $T$ come closer to complete spatial synchronization, the regime of chimera with intermittency is significantly transformed (Fig.~\ref{fig:zoo}(b)). In the incoherent part, the appearance of coherent subsets becomes more frequent and increased in space, at the same time its position does not stay in one place. This is especially clearly seen from the local order parameter and averaged phase velocity (3rd and 4th subpanels in Fig.~\ref{fig:zoo}(b)). This regime can be found in yellow regions of the map Fig.~\ref{fig:maps_pulse}(a).

The light-orange regions in the map Fig.~\ref{fig:maps_pulse}(a) correspond to a breathing chimera given in Fig.~\ref{fig:zoo}(c). The breathing chimera is characterized by the oscillating boundary between coherent and incoherent domains \cite{Bolotov2017}. This regime has been already observed in different systems \cite{Abrams2008, Ma2010, Kemeth2016}, mainly in ensembles of phase oscillators, but for the ring of autonomous FHN systems this regime have not been previously observed.

On the border between two synchronization tongues (Fig.~\ref{fig:maps_pulse}(a), dark-blue region), the another chimera may appear (Fig.~\ref{fig:zoo}(d)). This chimera contains two incoherent domains and corresponds to multi-chimera states~\cite{Omelchenko:2015vr}. The same chimera has been also found in a ring of coupled FHN systems in excitable regime \cite{Semenova:2020aa}. 

With a large amplitude of external influence, another interesting mode appears (see chimera* in Fig.~\ref{fig:zoo}(e)). This mode looks like an origin chimera when considering only snapshots or local-order parameter, but the mean phase velocity is the same for all oscillators, therefore it cannot be called as a chimera in its common form. Essentially, there is a spatial synchronization with a phase shift. For periodic-pulse exposure, chimera* is mainly observed for small period $T$ (see hatched regions on the map Fig.~\ref{fig:maps_pulse}), and there it can be observed from only specially prepared initial conditions of stabilized chimera. The random initial conditions lead to spatial synchronization with a phase shift without any clear domains. This regime of spatial synchronization with a phase shift is shown in light-blue color on the map Fig.~\ref{fig:maps_pulse}; the hatched areas overlaid on top of them indicate the appearance of chimera*.

All the above regimes have not been observed for the ring of autonomous FHN systems, therefore we can assume that these regimes are induced by external periodic-pulse influence. The question is: are these modes caused precisely by pulsed influence or by any periodic influence? In the next section we will consider the impact of harmonic influence on the same chimera, as well as the influence of purely positive harmonic influence.

\subsection{Regimes induced by harmonic influence}
In this section we consider the impact of harmonic influence on chimera state:
\begin{equation}\label{eq:harm_inf}
I(t) = A\cdot\sin w t,
\end{equation}
where $A$ and $w$ are amplitude and frequency of external force $I(t)$. 

\begin{figure}[b]
\includegraphics[width=\linewidth]{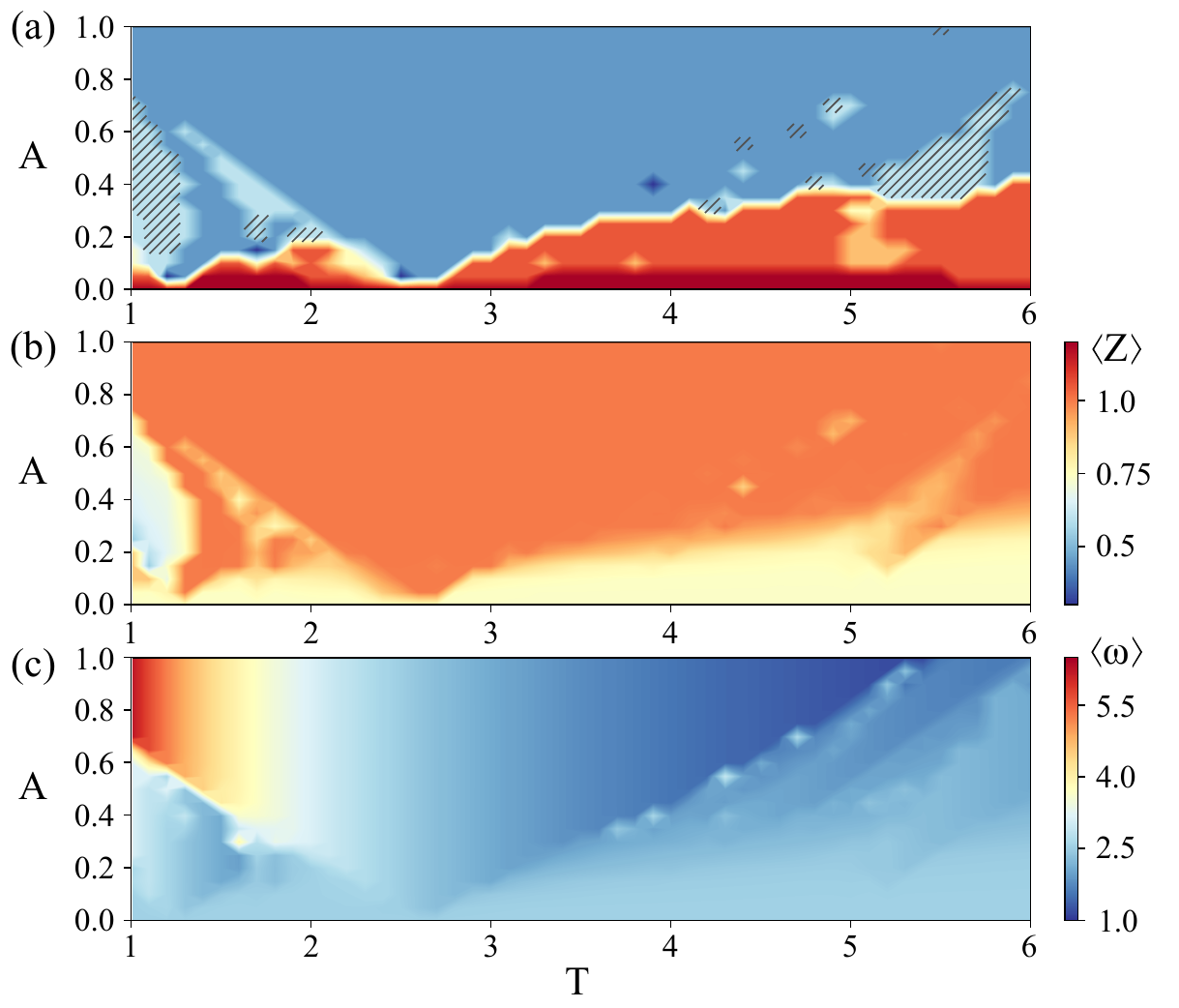}
\caption{\label{fig:maps_harm} Impact of harmonic external force on chimera state. Figure contains the map of regimes (a) in the parameter plane $(T, A)$, mean local order parameter (b) averaged over the integration time, and mean phase velocity (c) averaged in the same way. The rest parameters are the same as was in Fig.~\ref{fig:chimera_ome13}.}
\end{figure}

Figure \ref{fig:maps_harm}(a) contains the map of regimes in parameter plane of harmonic external influence (period and amplitude). This map of regimes was prepared in the same way as was before for Fig.~\ref{fig:maps_pulse}(a) with the same color scheme and designations. In addition, Fig.~\ref{fig:maps_harm} contains the values of averaged local order parameter (b) and mean phase velocity (c) in the same parameter plane $(T, A)$.

As was for periodic-pulse exposure, the large amplitude leads synchronization areas forming Arnold tongues on the map, but their position and size differ from the ones observed previously. The transition to complete synchronization occurs for smaller amplitude values. In between original chimera (red areas in Fig.~\ref{fig:maps_harm}(a)) and synchronization (middle-blue), chimera with intermittency (Fig.~\ref{fig:zoo}(a)) can be observed in dark-orange regions. Strong intermittency (yellow in Fig.~\ref{fig:maps_harm}(a)) occurs much less frequently than for pulsed exposure. The breathing chimera (light-orange regions) is also less common and occurs mainly at period $T\approx 5$. This value of period in pulsed exposure corresponded to a synchronization tongue. At the same time, there are several large areas of chimera* (Fig.~\ref{fig:zoo}(e)) shown as hatched area in Fig.~\ref{fig:maps_harm}(a), and this regime is observed not only for small $T$ values. However, for $T\approx 5.5$, chimera* can be obtained also from random initial conditions that was not in the case of periodic-pulse exposure.

\subsection{Regimes induced by positive harmonic influence}
Comparing the influences set by (\ref{eq:periodic_pulses}) and (\ref{eq:harm_inf}), one can see that the first $I(t)$ can only contain positive values, while the second covers both positive and negative values. Therefore, it would be correct to consider the harmonic influence shifted to the positive range:
\begin{equation}
I(t) = \frac{A}{2}\big(1 + \sin w t\big),
\end{equation}
where $A$ and $w$ are amplitude and frequency of external influence. 

\begin{figure}[b]
\includegraphics[width=\linewidth]{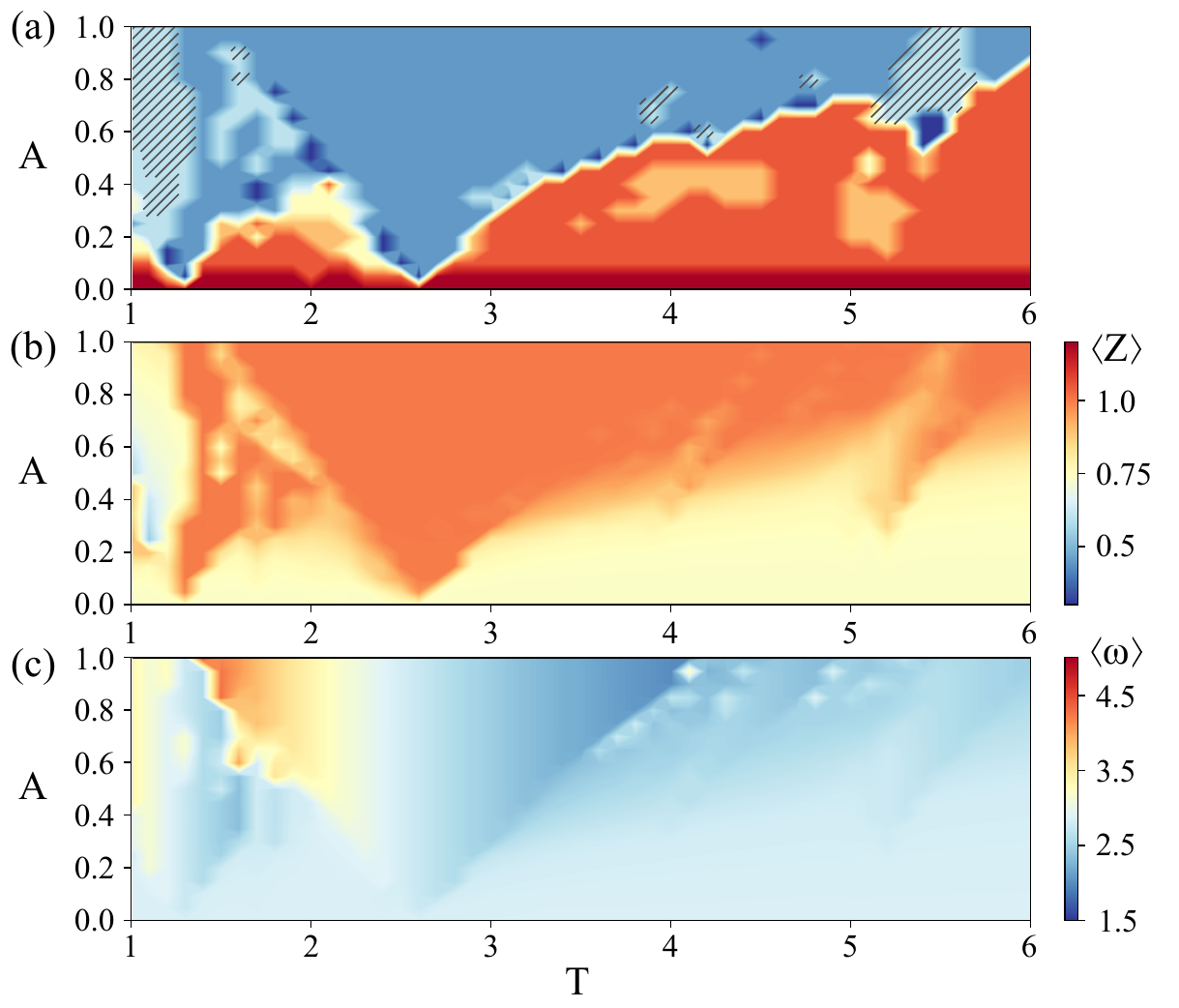}
\caption{\label{fig:maps_harm_pos} Impact of positive harmonic external force on chimera state. Figure contains the map of regimes (a) in the parameter plane $(T, A)$, mean local order parameter (b) averaged over the network and time, and mean phase velocity (c) averaged in the same way. The rest parameters are the same as was in Fig.~\ref{fig:chimera_ome13}.}
\end{figure}

Figure \ref{fig:maps_harm_pos} contains the map of regimes in the parameter plane of positive harmonic external influence (period and amplitude). This figure was prepared in the same way as was before for Fig.~\ref{fig:maps_pulse} and \ref{fig:maps_harm} with same color scheme and designations.

Comparing the maps prepared for harmonic and positive harmonic influences, it becomes clear that the areas of synchronization become more tiny and shifted to higher amplitude $A$ in the case of positive exposure, and this effect is similar to what was observed for pulse exposure. The values of $T$ corresponding to Arnold tongues are similar for both positive/negative and pure positive harmonic influence.

As the synchronization regions are reduced, the free space is covered by chimera with intermittency. However, strong intermittency is rarely observed, so we can assume that this mode is mainly caused by exposure with impulsive nature. Breathing chimera is mainly occurred by the positiveness of periodic influence, so it can be observed for both periodic pulse exposure and positive harmonic influence but not so pronounced for positive/negative harmonic influence. 

There are two areas of chimera* observed for small $T\approx1.1$ values and for $T\approx 5.5$. The first area becomes larger than for positive/negative harmonic influence from the previous section, while the second area remains approximately the same. For pulsed exposure, the first area of chimera* looks more like for positive harmonic influence, but the second area is completely absorbed by the synchronization tongue.

\section*{Conclusion}
In this paper, we have considered the impact of periodic pulse and harmonic exposure on the chimera state observed in the ring of nonlocally coupled FitzHugh-Nagumo systems. Here we have found that there are several different chimera states induced by this influence: breathing chimera, chimera with intermittency in the incoherent domain, travelling chimera with strong intermittency and finally chimera* with spatial synchronization and phase shift in the areas where incoherent domain is observed for original chimera. The last one regime is mainly induced by the periodic property of external influence, since it can be found much more often with simply periodic harmonic influence than for periodic pulses. And in general, the structure of the regime map for pulse exposure differs from the map for harmonic influence.

The pulsed exposure leads to an interesting regime of breathing chimera with periodically changing boundary between coherent and incoherent domains. This regime can be also found for harmonic influence, but in the case of periodic pulse action, several areas of this regime appear in the parameter plane $(T,A)$.

In this work, we showed the main special modes that can arise in the system due to pulse action. This work opens a series of works on studying the influence of pulsed action on a chimera. In subsequent works, we will show how noisy pulse influence affects the chimera state and what modes L{\'e}vy noise can lead to.

\begin{acknowledgments}
This work was supported by the Russian Science Foundation (project No. 23-72-10040) 
\end{acknowledgments}

\section*{Data Availability Statement}
The data that support the findings of this study are available from the corresponding author upon reasonable request.

\end{document}